# CRACKING THE LIU KEY EXCHANGE PROTOCOL IN ITS MOST SECURE STATE WITH LORENTZIAN SPECTRA[1]

LAZAR L. KISH[2], BRUCE ZHANG[2], AND LASZLO B. KISH[3]

*Department of Electrical and Computer Engineering, Texas A&M University, College Station, TX 77843-3128, USA*

We have found a security risk in the Liu's cypher based on random signals and feedback, when it utilizes a large class of noises for communication in its most secure state, the steady state. For the vulnerability to exist, the noise must have a spectrum which can be transformed to white-like noise by linear filtering. For the cracking, we utilize the natural properties of power density spectra and autocorrelation functions. We introduce and demonstrate the method for Lorentzian spectra. Some of the implications of the results concern the transient operation during changing bits, where the modulation products of noise cannot be band-limited therefore the cypher is vulnerable. We propose the application of line filters to provide a proper spectral shape and to improve the security.

## 1. Introduction

Recently, there has been an intensifying development in the field of unconditionally secure communication via separated classical physical systems [1-4]. They were originally inspired by the Kirchhoff-loop-Johnson-(like)-noise key exchange protocol (KLJN-cypher) [5-16] which however contains wired parties to provide a single, integrated physical system (Kirchhoff-loop) consisting of Alice's and Bob's communicators at the specifically selected low operational frequencies. The security of the idealized KLJN cypher is protected by the second law of thermodynamics, that is, by the impossibility of a perpetual motion machine of the second kind.

In two very recent papers [1,2], Liu has introduced and tested a new, very interesting type of secure key exchange protocol (Liu-cypher). If it is unconditionally secure, as claimed, it has the potential to revolutionize secure communication.

The particularly interesting property of the Liu cypher [1,2] stems from the fact that it is a classical physical system, just like the KLJN-cypher, however it is based on a completely separated pair of physical systems, which are sending only numbers to each other, even through email or mail. If the Liu cypher is indeed secure then it makes all the other secure communicators, RSA, quantum, KLJN, etc, obsolete, complicated, and unnecessary. On the other hand, no physical law has been identified as the protection of its security.

---

[1] .http://arxiv.org/abs/0910.0665
[2] On leave from A&M Consolidated High School, 1801 Harvey Mitchell Pkwy, College Station, TX, USA
[3] Corresponding author.

Note that communicators unconditionally secure at the conceptual level can never be absolutely secure at practical applications due to non-idealities; and this statement is valid also for quantum communicators. However, if Alice and Bob can exchange more key bits than the information accessible for Eve via eavesdropping, privacy amplification algorithms will allow an arbitrarily enhancement of the actual security by generating a short key with enhanced security from the original longer key with greater information leak.

Therefore, the essential question of cracking any secure physical communicator is as follows: Can Alice and Bob exchange more information about the key than the information Eve can extract during the key exchange process? If the answer for the Liu cypher is yes then it can be made arbitrarily secure. However, if the answer is no then the cypher has zero security.

## 2. The Liu-cypher based on feedback and noise

Dr. Liu's has made several attempts to extract the essence of the KLJN cypher and implement it in new systems without thermal noise and Kirchhoff-law aspects. The first attempt was an interesting circulator-based model [6] which was criticized, further developed, and finally cracked in [7] by a circulator-based man-in-the-middle attack.

As we have already mentioned and we want to further emphasize, the newest, very interesting development [1,2], the Liu cypher, does not require a physical system or physical law, at all. Even two computers communicating via email or, in principle, two people communicating with regular mail can use it, if speed is not a problem. And, if the method works, it is automatically protected even against the man-in-the-middle attack by broadcasting the signals by Alice and Bob. (Note, broadcasting is different from authentication, which was a mistake in [1,2]; this is a small but important correction.)

The important question is if the Liu-cypher [1,2] can generate and share an unconditionally secure key by just sending numbers back and forward between Alice and Bob. Philosophically, it is very difficult to imagine *unconditional* security (even at the conceptual level) in such a way, though such generalized attempts have been already made, but with no success [17-19].

The protocol of the Liu cypher [1,2] is as follows (see Figure 1). Alice and Bob choose their own small reflection coefficient $\alpha$ and $\beta$ with random (secret) signs and publicly known uniform absolute value $|\alpha| = |\beta| = \gamma \ll 1$. The secret arrangement of signs stays valid for the whole clock period. Then, see Figure 1, Alice and Bob reflect the incoming signals $X_{BA}$ and $X_{AB}$, according to their own reflection coefficients, and also add their own secret Gaussian random noises $V_A(t)$ and $V_B(t)$. The effective values of noise amplitudes and the noise spectra are equal and publicly known. When they happen to select a reflection coefficient with opposite signs, $\alpha = -\beta$, a secure bit is generated and exchanged during the clock period. For the equations [1,2], see Figure 1.

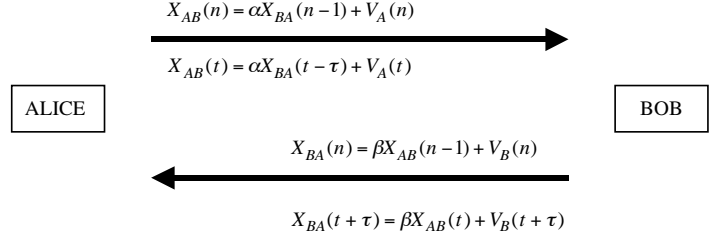

**Figure 1.** The protocol of the Liu cypher [1,2]. Continuum time and discrete time versions are given. The duration of signal roundtrip (propagation+processing) is $2\tau$, or two time-steps, respectively.

The parameters are chosen so that the shortest time constant is $\tau$ which is half of the signal roundtrip time (propagation+processing). The noises are chosen to have such a long correlation time $\tau_c$ ($\gg \tau$) (it means a small bandwidth spreading from zero frequency up to $B_n$) that they can be considered *static* during the signal roundtrip time. Under this condition, the system is converging to a geometrical series (see [1,2]) with power exponent $\gamma^2$ and coefficients dictated by the linear combination of the actual amplitudes of the noises $V_A$ and $V_B$. The longest time-parameter is the clock period $\tau_{bit}$, which is long enough to include many correlation times of the noise, in order to have a good statistics, when the noises, signals, and their combinations are time averaged. In conclusion:

$$\tau \ll \tau_c \approx \frac{1}{B_n} \ll \tau_{bit} \tag{1}$$

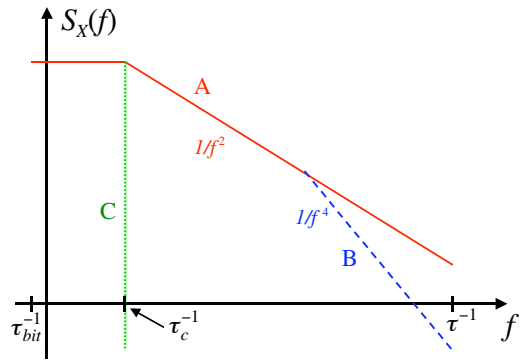

**Figure 2.** Examples for different noise spectra with the same -3 dB bandwidths and similar noise bandwidths. A: Lorentzian, white noise filtered by a first-order low-pass filter; B: white noise filtered by a second-order low-pass filter; C: infinitely steep cut-off by digital filtering. Only type C is secure (see Section 3) but it cannot be reached during normal operation due to transients (see Section 5).

Alice and Bob extract the sign of the reflection coefficient of the other side by cross-correlating the returning signal with their own noise [1,2]. The sign of the cross-correlation coefficient between the local noise amplitude at time $t-\tau$ and the returning signal amplitude at time $t$ is obviously the same as the sign of the reflection coefficient of the other side [1,2].

In the steady-state, when the geometrical series characterizing the system had practically reached its actual stationary value (remember the noise was virtually static during the time scale of the convergence), the Liu cypher was claimed absolutely secure [1,2]. However, it was recognized [1] that during the convergence to the steady state, for example during the initial transient at the beginning of the clock period, the cypher was leaking information. Various tricks were proposed to fix these weaknesses [1].

In the present paper, we show an attack against the Liu cypher in its state where it was believed to be absolutely secure: in its steady state. The results indicate that the shape of spectrum of the noise is an important parameter not only its bandwidth as it was originally believed [1,2].

### 3. Cracking the security of Liu-protocol for Lorentzian noises

We will see that to break the protocol with our presently proposed method, it is essential that the correlation time of the observed quantities is in the order of $\tau$. This may look paradoxical because such claim is obviously not valid for the added noises (and resulting signals which have the same $\tau_c$), see Eq. 1. To provide the necessary condition for Lorentzian and similar noises with high-frequency tails a pre-processing will be needed.

The most important reason why our general method of cracking works is the consequence of the well-known fact that the autocorrelation time of an idealistic white noise is zero.

First, we crack the protocol for the simplest case of added noises $V_A(t)$ and $V_B(t)$ when they have a Lorentzian spectral shape $\left[1+(f/f_c)^2\right]^{-1}$, where the cut-off frequency and the correlation time are interrelated as $f_c = \tau_c^{-1}$. The practical importance of a Lorentzian is that it results from white noise by a simplest first-order low-pass filter with a single pole $\left[1+j\dfrac{f}{f_c}\right]^{-1}$. It is important to remember that the spectra of the noises are publicly known quantities.

For Lorentzian noise, the simplest way of cracking is as follows. The spectrum of the time derivative of a noise with spectrum $S(f)$ results in $4\pi^2 f^2 S(f)$ spectrum of the velocity and that will shorten the correlation time of the noise to the required regime. The reason is that in the case of the Liu cypher the new spectrum becomes $f^2$-noise in a narrow frequency range, at low frequencies, and white noise in the wide range at high frequencies. Therefore, the correlation time of that white-like noise will approach the shortest observable time constant in the system which is $\tau$. Thus, Eve can cross-correlate the velocity of their sent-out signal with the returning one and that will provide the most efficient way to extract the reflection coefficients. Below, we show how to extract the sign of the reflection coefficient at Alice. The sign of the following velocity crosscorrelation will tell Eve the bit of Alice because its sign will be equal to the sign of $\alpha$ :

$$\Phi_{AB}(t)\Phi_{BA}(t-\tau) = \left[\alpha\Phi_{BA}(t-\tau) + W_A(t)\right]\Phi_{BA}(t-\tau) =$$
$$= \alpha\Phi_{BA}^2(t-\tau) + \beta W_A(t)W_A(t-2\tau) + \alpha\beta\Phi_{BA}(t-\tau)\Phi_{BA}(t-3\tau) \quad (2)$$

where the $\Phi_{i,j}(t)$ quantities are the time derivatives of the corresponding $X_{ij}(t)$ signals and the $W_k(t)$ quantities are the time derivatives of the corresponding $V_k(t)$ noise secrets. The coefficient of the last term at the right-hand-side of Eq. 2 is a small quantity ($\alpha^2\beta$) proportional to $\gamma^3$ which can be dropped. After time averaging, we get:

$$\left\langle \Phi_{AB}(t)\Phi_{BA}(t-\tau) \right\rangle_t = \alpha\Phi_0^2 + \beta\Gamma_{W_A}(2\tau) \approx \alpha\Phi_0^2 , \quad (3)$$

where $\Phi_0^2$ (> 0) is the mean-square signal velocity, $\Gamma_{W_A}(2\tau)$ is the autocorrelation function (with $2\tau$ time-shift) of $W_A(t)$. As we have pointed out above, the time derivative of the Lorentzian noise will have zero autocorrelation function for time shifts $\tau$ or greater, thus $\Gamma_{W_A}(2\tau) = 0$. Therefore, the sign resulting from Eq. 3 will show the sign of $\alpha$.

Computer simulations for the case of Lorentzian noise generated from damped Brownian motion and $\gamma = 0.2$ show, see Table 1, that Eq. 3 will crack the cypher with excellent success rate, greater than 99.999%, within a single correlation time of the noise when the clock duration $\tau = 1000$ steps. For clock duration $\tau = 100$ steps, which is the lower limit of reasonable $\tau_c$ correlation times, the same accuracy is obtained within 5 correlation times of the noises. These success rates and speeds are much greater than those of indicated between Alice and Bob in [1,2], and this situation is a convincing fact about the efficiency of the cracking method of Eve.

At this point, we could conclude the paper and stating that the Liu cypher was cracked for Lorentzian noise. However, Alice and Bob can also learn about the advantage of using velocity correlation functions and they can enhance their original protocol by using their $W_A(t)$ and $W_B(t)$ noise velocities to do the crosscorrelations instead of the $V_A(t)$ and $V_B(t)$ noise amplitudes originally proposed by Liu [1,2]. Thus, without improving the Liu cypher, by utilizing the new idea of velocity correlations and comparing the improved cypher with Eve's cracking protocol, it is unclear how much security actually remains in the new situation. It is because Alice and Bob may use much shorter clock cycles with the enhanced cypher thus they may reduce the effectiveness of Eve's method. The improved protocol will be:

*Cracking the Liu protocol of secure key exchange*

$$\Phi_{AB}(t)W_B(t-\tau) = \left[\alpha W_B(t-\tau) + W_A(t)\right]W_B(t-\tau) = \\ = \alpha W_B^2(t-\tau) + W_A(t)W_B(t-\tau) \qquad (4)$$

After time averaging:

$$\left\langle \Phi_{AB}(t)W_B(t-\tau) \right\rangle_t = \alpha \left\langle W_B^2(t-\tau) \right\rangle_t = \alpha W_0^2 \qquad (5)$$

where $W_0^2$ is the mean square of $W_B$.

*The remaining but ultimate question is if Eq. 5 is more efficient than Eq. 3. If yes, the security can be saved by privacy amplification.*

However the operation described by Eq. 5 is less accurate than using Eve's eavesdropping protocol shown above because the terms resulting the DC components in Eqs. 2 and 4 (see the middle section of the equations) are related as:

$$\Phi_0^2 = \frac{1+\gamma^2}{\left(1-\gamma^2\right)^2} W_0^2 \qquad (6)$$

see the results [1,2]. On the other hand, the terms representing the noise (to be averaged out) in Eqs. 2 and 4 (see the middle section of the equations) are related as:

$$\sqrt{\left\langle \left[W_B(t)\Phi_A(t-\tau)\right]^2 \right\rangle} \approx \frac{\sqrt{\left\langle \left[W_B(t)W_A(t-\tau)\right]^2 \right\rangle}}{1-\gamma^2} \qquad (7)$$

Thus, the signal-to-noise ratio of Eve's method is $(1+\gamma^2)/(1-\gamma^2) > 1$ times greater than that of Alice's and Bob's new method. This difference results in a greater error rate for Eq. 5. In conclusion, Alice and Bob must use Eve's method, Eq. 3, to obtain the highest speed and the lowest error rate.

| $\tau_{bit}$ (steps) | Eve (Eq. 3) $\tau_c = 100$ (steps) | Eve (Eq. 3) $\tau_c = 1000$ (steps) | Alice/Bob (Eq. 5) (Improved Liu) $\tau_c = 100$ (steps) | Alice/Bob (Eq. 5) (Improved Liu) $\tau_c = 1000$ (steps) |
|---|---|---|---|---|
| 50 | 85.0% | 84.2% | 73.5% | 71.0% |
| 100 | 95.6% | 95.3% | 88.4% | 83.9% |
| 200 | 99.5% | 99.5% | 97.8% | 93% |
| 500 | >99.999% | >99.99% | >99.9% | 98% |
| 1000 |  | >99.999% |  | >99.9% |

**Table 1.** Computer simulation results for Lorentzian noise with Eve's cracking method (Eq. 3) and the enhanced Liu cypher (Eq. 5), at two different correlation times of the secret noises.

Table 1 shows computer simulation results comparing Eve's cracking method (Eq. 3) and the enhanced Liu cypher (Eq. 5) with Lorentzian noise spectrum, at two different correlation times of the secret noises. It can be seen that even though the Liu cypher gets progressively enhanced compared to the original version [1,2], it still performs weaker than Eve's method. Thus Alice and Bob must use Eve's method and that means zero security.

**4. Other types of noise spectra**

The generalization of the cracking principle described in Section 3 for Lorentzian noises is simple. It is based on linearly transforming the observed signal to another quantity with white-like noise characteristics. Let us suppose that the (publicly known) spectrum of the noises and the signals is $S_X(f)$. Then the observed signals $X_{AB}$ and $X_{BA}$ must be sent through a linear filter with amplification $A(f)$ :

$$|A(f)| = \sqrt{\frac{1}{S_X(f)}} \qquad (8)$$

Then after the filtering, the new quantities $Y_{AB}$ and $Y_{BA}$ will have a white-like noise spectrum $S_Y(f)$ in the whole frequency band of operation:

$$S_Y(f) = |A(f)|^2 S_X(f) = 1 \qquad (9)$$

*Due to the linearity of these operations*, the rest of the cracking operation and the other considerations in Section 3, remain the same, as above.

*a.* The $Y_{AB}$ and $Y_{BA}$ quantities must be used with timing corresponding to the timing of $X_{AB}$ and $X_{BA}$.

*b.* The sign of $\langle Y_{AB}(t)Y_{BA}(t-\tau)\rangle_t$ will provide the sign of $\alpha$.

*c.* All the other considerations, comparisons, and conclusions in Section 3 remain the same because the white noise case provides the best statistics within the shortest duration. Thus, the best method Alice and Bob can use with utilizing their noises is the situation when they send their own noise and the observed signals through the same filtering process. However, Eve's method described in Section 3 stays more efficient.

It is important to note that, for the cracking method to work, the noise spectrum must have non-zero value up to $\tau^{-1}$. Thus a steep cut-off indicated by C in Figure 2 is immune against this type of attack. However, noise made from white noise filtered by classical first, second, etc orders low-pass filters are vulnerable. Note, Liu has mentioned type C spectra in [2] however the goal was not to secure security but the have a convenient autocorrelation function. It has been realized first in the present work that not only the noise-bandwidth but also the spectral shape matters. About practical implications, see the next section.

## 5. Implication: relevant type of vulnerability during transient response

Because this attack took place in the working mode where the cypher was believed to be absolutely secure, a particular attention must be given for this type of attack in the transient period. This is particularly important because, even if the noise had type C shape, which remains secure in the steady-state mode, in the transient regime this may produce combination spectra of A, B, or similar spectra with non-zero amplitude. This fact implies the necessity to use line filters to secure the type C shape, and this requirement has not been recognized earlier. But even with idealistic line filters, further security tests are necessary during normal (transient) key exchange operations by utilizing the new type of attack described here, in order to assess the practical level of the remaining security.

## 6. Conclusion

A substantial vulnerability for Lorentzian and similar noises have been discovered in the Liu cypher. The vulnerability presents even in the steady state where the cypher was supposed to be unconditionally secure. In the steady state, the vulnerability can be avoided by idealistic, infinitely steep cut-off at the high-frequency end of the noise spectrum. However, staying in the steady state means zero information channel capacity and that means no communication. As soon as real communication begins, and bits are changed, the originally idealistic band-limited noise will produce cross-modulation components with the transients and that means infinite high-frequency tails in the signal spectrum.

Thus perhaps the most important practical implication of the present results is the minimal need of applying line filters during normal (transient) operation.

### Acknowledgements

We appreciate discussions with Dr. Pao-Lo Liu to whom we congratulate for his scheme and the challenge it resulted. We appreciate useful comments from Dr. Mark Dykman and Dr. Jacob Scheuer, too.